\documentclass[11pt]{cernrep}
\usepackage{graphicx,amsmath}
\begin{document}

\title{Study of color superconductivity with 
Ginzburg-Landau effective action on the lattice%
\thanks{$\;$ Talk presented by M.~Ohtani
at the workshop on Extreme QCD (Swansea, Aug. 2-5, 2005).}}

\author{Sanatan Digal$^{1,2,3}$, Tetsuo Hatsuda$^1$ and Munehisa Ohtani$^2$}

\institute{
$^1$ Department of Physics, University of Tokyo, Tokyo 113-0033, Japan \\
$^2$ Radiation Laboratory, RIKEN,
2-1 Hirosawa, Wako, Saitama 351-0198, Japan\\
$^3$ Institute of Mathematical Sciences, Taramani, Chennai 600 113, India.
}

\maketitle
\vspace*{-13.5em}
\begin{flushright}
{\small \tt TKYNT-05-29, IMSc/2005/11/26}
\end{flushright}
\vspace*{10.15em}
\begin{abstract}
We study thermal phase transitions of color superconductivity by the lattice
simulations of the Ginzburg-Landau (GL) effective theory. The theory is
 equivalent to the SU$_{\rm f}$(3) $\times$ SU$_{\rm c}$(3) 
Higgs model coupled to SU$_{\rm c}$(3) color gauge fields. 
From the eigenvalues of a 3$\times$3  gauge-invariant diquark composite,
a clear distinction between the 2-flavor color superconductivity (2SC) and the
color flavor locking (CFL) phase is made in a gauge invariant manner.
The thermal transitions  between the normal phase and the superconducting phases
are found to be first-order due to thermal gluons. 
The phase structure in the coupling-constant space is numerically explored and
three patterns of phase transition, i.e., normal$\rightarrow$2SC, 
normal$\rightarrow$CFL and normal$\rightarrow$2SC$\rightarrow$CFL, 
are found in the chiral limit. These results agree qualitatively
with the weak-coupling analysis of the GL theory.
\end{abstract}

\section{Introduction}

 In dense and cold quark matter,  
 color superconducting phase is expected to be realized due to
  diquark pairing (see the reviews, \cite{Al}).
 From the analysis of the 
 pairing interaction, the most attractive channel is found to be 
in the color anti-triplet and $J^P=0^+$ channel, which
 leads to the  diquark field,
$\varPhi_{cf} =  \epsilon_{abc} \epsilon_{fjk}
\langle q^{j}\!\!_{a} {\rm C}\gamma_5 q^{k}\!\!_{b} \rangle$,
 as the most relevant quantity to describe color superconductivity.
 A three-dimensional Ginzburg-Landau (GL)   effective
action  written in terms of  $\varPhi_{cf}$ was
 proposed \cite{IB} to study the thermal phase transition
 of the color superconductivity. Then, it was realized  that
 the thermal gluons (treated in weak-coupling approximation)
 turn the second-order transition in the mean-field theory
  to the first order transition \cite{Mat,Ren}.
 Such a weak-coupling study, however, is valid only in extremely
 high density  because the gauge coupling grows for small
 baryon density. Furthermore, thermal fluctuations of the 
diquark field become also important
 as density decreases \cite{Mat}.

 Motivated by this observation,
 we carry out  Monte-Carlo  simulations of the GL effective action
 discretized on the lattice.
Although the simulation of the GL effective action 
is not {\it ab initio} calculation and is applicable only
 near the thermal phase boundary,
  it has an advantage of making non-perturbative calculation 
 and simultaneously of  avoiding the sign problem 
 which is a serious obstacle in direct lattice QCD simulation at
finite density.
Since the GL action has color SU$_{\rm c}$(3) gauge invariance and 
 the diquark field $\varPhi_{cf}$ is a gauge variant
  object, it is not a trivial task to identify
  the color superconducting phase in the lattice approach.
 We will see below that the eigenvalues of a gauge invariant
 composite $\sum_x\varPhi^\dag_x\varPhi_x$
 are useful to identify different phases. ($x$ denotes lattice sites.)
Also, analyzing the hysteresis loops of such gauge invariant observables,
 we find the first-order transitions among these phases.
 
\section{The Ginzburg-Landau effective action}

 The 3-dimensional Ginzburg-Landau (GL) action  in the chiral limit \cite{IB},
which is expected to be universal around  the thermal phase boundary, reads
$S_{\rm GL} = \int d^3x {\cal L}$ with 
\begin{equation}
{\cal L}=\frac{m^2}{2}{\rm Tr}(\varPhi^\dag\varPhi)
+\frac{\lambda_1}{4}\left[{\rm Tr}(\varPhi^\dag\varPhi)\right]^2 
+\frac{\lambda_2}{4} {\rm Tr}(\varPhi^\dag\varPhi)^2 + 
\frac{\kappa}{2}{\rm Tr}|D_i\varPhi|^2+\frac{1}{4}F_{ij}^aF^{aij},
\label{GL}
\end{equation}
where we have assumed three massless quarks.
Since the diquark field is anti-triplet in color space,
 the covariant derivative is defined by
$D_i\varPhi\equiv(\partial_i-i g T^{a *}A_i^a)\varPhi$.
Effects of  temperature and baryon chemical potential are incorporated
in the coupling constants ($m$, $\lambda_{1,2}$ and $\kappa$).
The action has global SU$_{\rm f}(3)$ flavor symmetry and
local SU$_{\rm c}(3)$ gauge symmetry.
 
 Since $S_{\rm GL}$ is real, sign problem does not occur.
  In our actual simulation, 
 we extend Eq.(\ref{GL}) slightly to {\sl four dimension} with two  
 temporal slices and  periodic boundary condition. We do so to define
 the Polyakov loop which can be used as a measure to see if
  the system is in the  
  deconfinement phase. This extended 4-d action  is expected to 
 give qualitatively the same results as the 3-d action,  although
  the parameters in the former and those in the latter
   ($m$, $\lambda_{1,2}$ and $\kappa$)
 are related in a non-trivial way by  the renormalization from
  the temporal field-fluctuations.
  The GL action in Eq.(\ref{GL})
  is similar to the gauged Higgs model which is 
  studied in the context of the electroweak phase transition.
 Note, however, that we have two independent coupling 
 $\lambda_1$ and $\lambda_2$ unlike the case of the SU(2) Higgs model.

\section{Weak-coupling analysis}

\begin{figure}[t]
\begin{center}
\raisebox{15em}{a)} \hspace{-.9em}
\includegraphics[width=6.3cm]{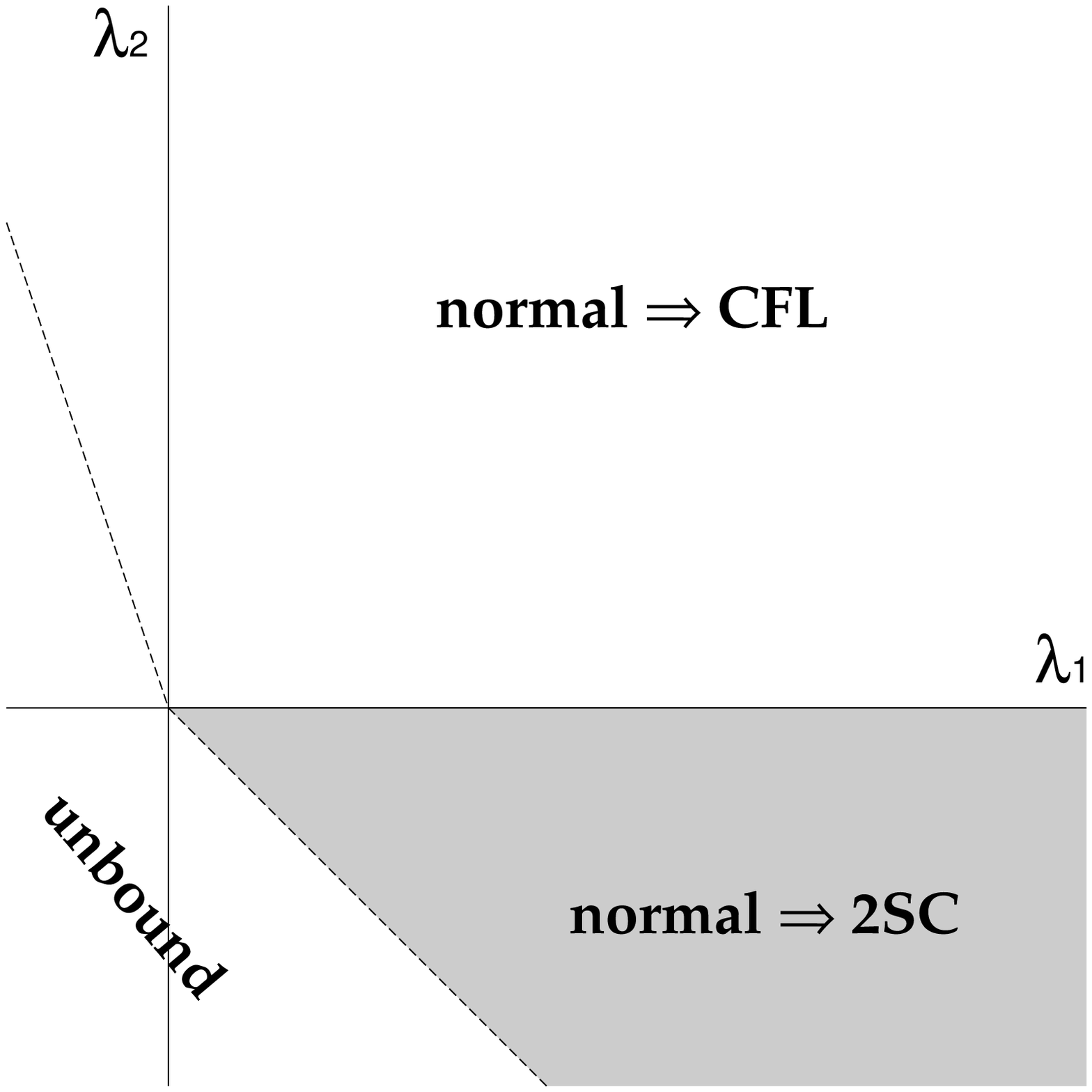} \label{mf}
\hspace{2.5em}
\raisebox{15em}{b)} \hspace{-.9em}
\includegraphics[width=6.3cm]{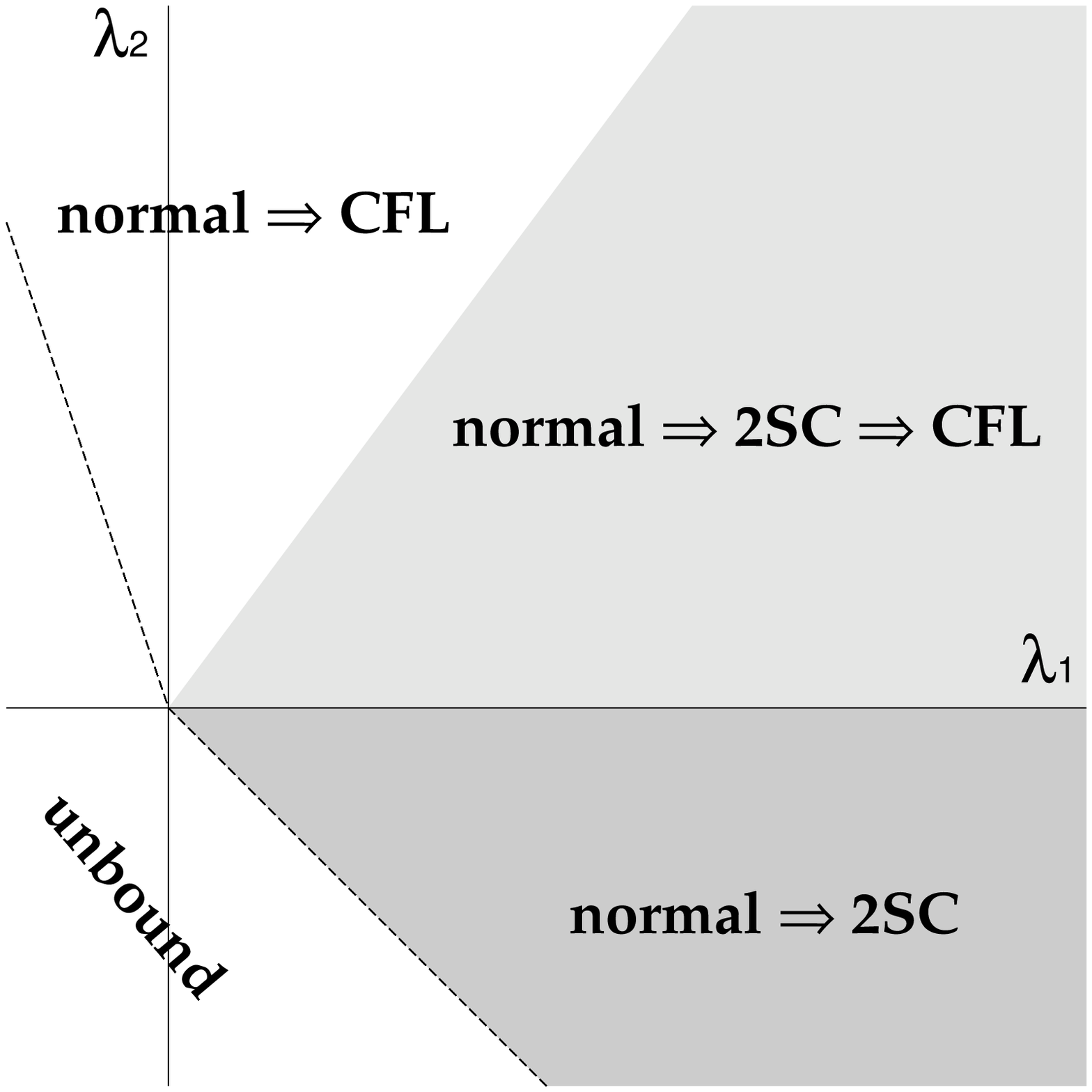}  \label{pt}
\vskip -0.3cm
\caption{a) Phase diagram of the 3-d GL theory in the coupling constant space 
 obtained by the mean-field theory \cite{IB}. The second order
  phase transitions are obtained.
b) Phase diagram obtained by
 taking into account the gluon fluctuations in a weak-coupling approximation.
  The phase boundary is shifted and the transition becomes first order.}
\vskip -0.3cm
\end{center}
\end{figure}

 The mean field approximation to the 3-d GL action 
 leads  to the second order transition between normal phase and
the color superconducting phase \cite{IB}. 
Fig.1a shows the phase diagram in the ($\lambda_1$, $\lambda_2$)-plane
 in the mean field approximation. 
 Whether 2SC or CFL is realized in color superconductor in the chiral limit
 depends on the sign of $\lambda_2$, and the phase boundary
 is located  on $\lambda_2=0$.

 For weak gauge-coupling, 
 it has been shown that the thermal fluctuation of gauge field 
dominates over  the diquark fluctuation, and can be incorporated 
as a Gaussian fluctuation around the mean-field \cite{Mat,Ren}.
 As a result, the phase transition turns into 
{\sl first order}, which is similar to the situation in Type I
  metallic superconductor \cite{HLM}.
Shown in Fig.1b is a phase diagram in the ($\lambda_1,\lambda_2$)-plane
obtained by using the weak coupling approximation along the line of ref.\cite{Mat}.
 Phase boundary between 2SC and CFL in Fig.1a
 is shifted towards the positive $\lambda_2$ region
 by the gluon fluctuation. Furthermore, with certain parameter region,
 we observe successive transitions, normal$\rightarrow$2SC$\rightarrow$CFL,
  as temperature decreases from above.
 Since  $\lambda_1=\lambda_2$ is realized in the weak-coupling limit,
 such  successive transitions may take place in high density QCD
  in the chiral limit. This possibility was first noted 
  in \cite{JPS}.  

\section{Lattice simulations}

 Let us discretize the 4-d version of the effective action, Eq.(\ref{GL}),
   and rescale the field and couplings as  
$\varPhi\to \sqrt{2\alpha}\varPhi /a$,
$\lambda_i\to {\bar \lambda}_i/\alpha^2$,
 $m^2\to (1-2\bar{\lambda}_1-2{\bar \lambda}_2-8\bar{\kappa})/\alpha a^2$,
$\kappa\to\bar{\kappa}/\alpha$ ($a$ being the lattice constant).
 Then, we reach the following lattice action
\begin{align}
   S= \sum_x & \left\{ {\rm Tr}(\varPhi^\dag_x\varPhi_x)+(\bar{\lambda}_1+\bar{\lambda}_2)
\left({\rm Tr}(\varPhi^\dag_x\varPhi_x)-1\right)^2+
\bar{\lambda}_2\left({\rm Tr}(\varPhi^\dag_x\varPhi_x)^2-
\left[{\rm Tr}(\varPhi^\dag_x\varPhi_x)\right]^2\right) \right. \nonumber\\ 
& \ \ 
-2{\bar \kappa}\sum_\mu {\rm Re\ Tr}(\varPhi^\dag_{x+\mu}U_\mu^*\varPhi_x)
+S_{\rm g}(\beta) \Big\}, \label{act}
 \end{align}
where $S_{\rm g}$ is the standard plaquette action  with $\beta \equiv 6/g^2$.
 To ensure that the system is in the deconfined phase, we keep
 $\beta$ around 5.1 according to the critical value obtained in
the pure Yang-Mills case with $N_\tau=2$. 
Taking  several sets of the quartic 
couplings $\bar{\lambda}_{1,2}$, we make simulations by scanning the parameter
$\bar{\kappa}$. In this article, we report the results obtained in the
simulation with a lattice volume of $2\times 12^3$.
 We note that the  $2\times 8^3$ lattice gives qualitatively the same results.
 We use the pseudo heat-bath method to update the gauge-link
 variables and generalize the efficient update-algorithm of 
SU(2) Higgs-field \cite{b} to our case.

\begin{figure}[t]
\begin{center}
\raisebox{13em}{a)} \hspace{-.9em}
\includegraphics[width=7.4cm, clip=yes]{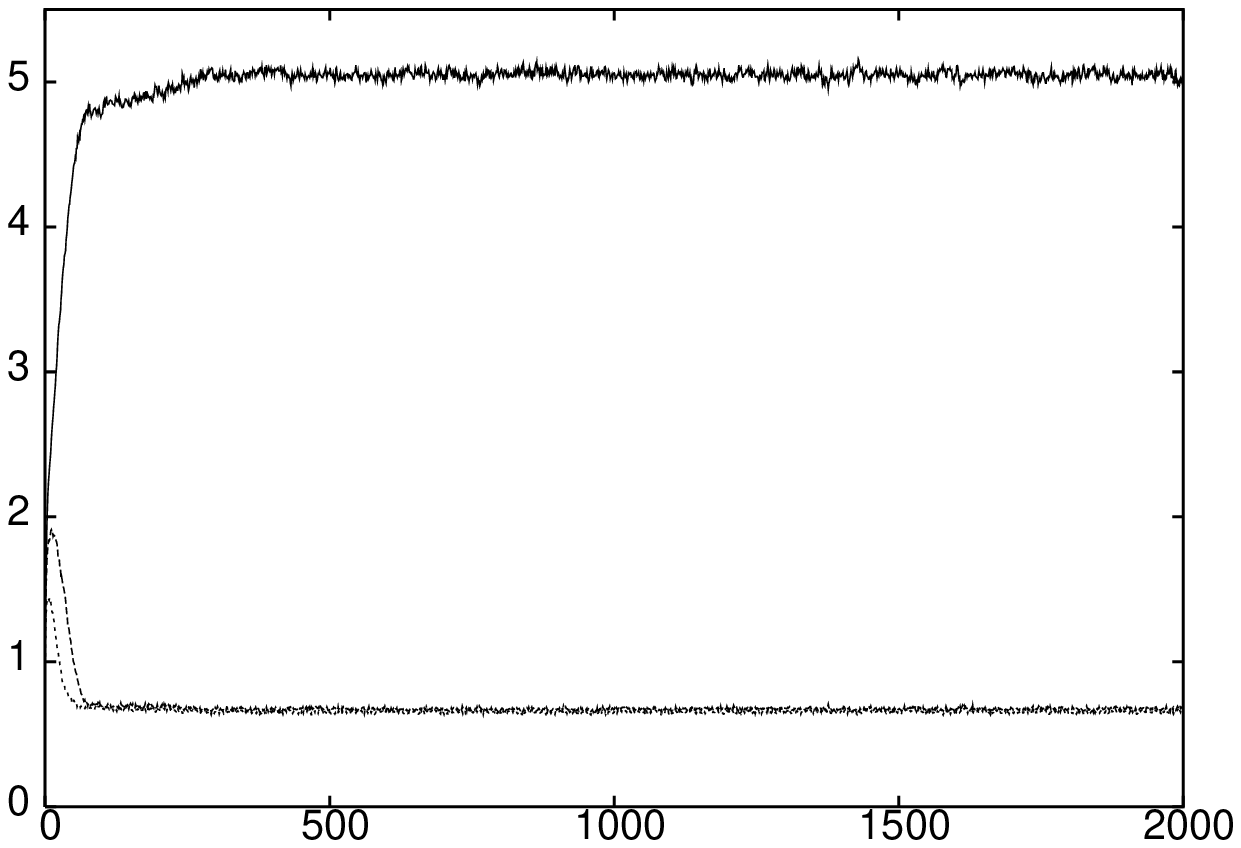} \label{ev2}
\hspace{.5em}
\raisebox{13em}{b)} \hspace{-.6em}
\includegraphics[width=7.4cm,clip=yes]{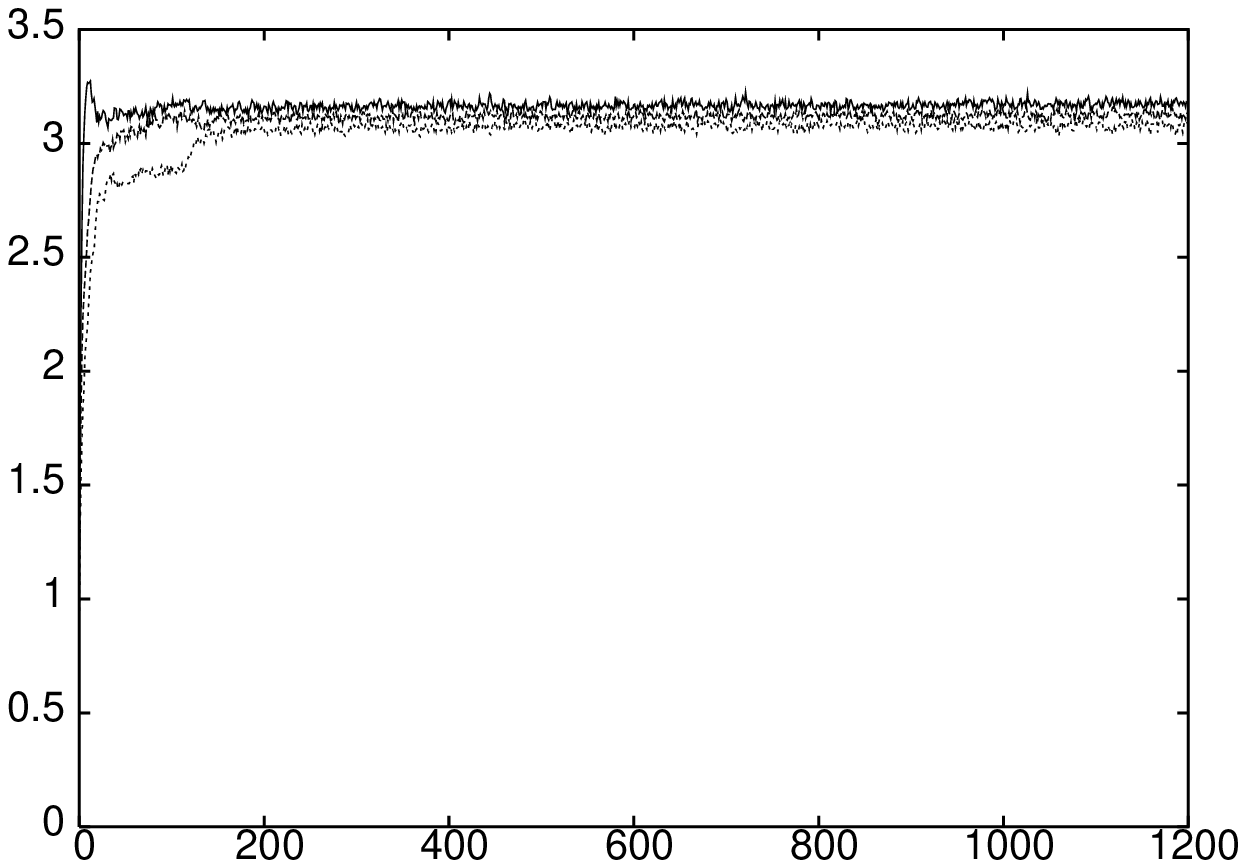} \label{ev1}
\vskip -0.3cm
\caption{Eigenvalues of $\sum_x\varPhi^\dag_x\varPhi_x$ as a function of 
update steps.  
$(\bar\lambda_1,\bar\lambda_2)=(0.03,0.01)$, $\bar\kappa=0.34$,
$\beta=5.1$ for the left panel a) and 
$(\bar\lambda_1,\bar\lambda_2)=(0.03,0.03)$, $\bar\kappa=0.43$,
 $\beta=5.1$ for b).
The isolated large eigenvalue associated with
 two degenerate small eigenvalues a)
 corresponds to 2SC phase while 
three degenerate eigenvalues after thermalization b) indicate CFL phase.}
\vskip -0.3cm
\end{center}
\end{figure}

 The color superconducting phase and the normal phase are distinguished
  by looking at the gauge invariant operator
$\sum_x{\rm Tr}\ \varPhi^\dag_x\varPhi_x$. Although the field fluctuations
give non-vanishing expectation value of this operator even in the normal phase, 
we can observe discontinuous changes of the expectation value as $\bar\kappa$ changes.
In addition to this, we measure other gauge-invariant operators 
such as  ${\rm Tr}|D_\mu \varPhi|^2, S_{\rm g}$ and the Polyakov loop to
confirm that the discontinuous change takes place 
 for all these operators at the same $\bar\kappa$.
(Note that this discontinuous change of the Polyakov loop does not
mean deconfinement transition.)
 This constitutes an evidence of the first order phase transition. 

 Moreover, we can differentiate different color superconducting states
(2SC, CFL or something else)
by diagonalizing the 3$\times$3 matrix
$\sum_x\varPhi^\dag_x\varPhi_x$ in the flavor space
and extracting its eigenvalues. 
 As shown in Fig.2a, we observe
   one large eigenvalue with two degenerate small
eigenvalues in a particular set of $(\bar{\lambda}_1,\bar{\lambda}_2$).
  In this case, 
we regard the state as 2SC in which the quark pairing 
 takes place in one specific color-flavor channel.
 We also observe three degenerate eigenvalues as shown in Fig.2b
with another parameter set, which corresponds to the CFL phase where
all channels are involved equally to the pairing. Here we emphasize that
this is a {\sl gauge invariant} way of identifying different phases 
 in color superconductivity.
As we scan the parameter $\bar{\kappa}$, we find
rapid changes of the eigenvalues of   $\sum_x\varPhi^\dag_x\varPhi_x$
 at critical value $\bar{\kappa}_{\rm c}$.
When we simulate around this point by increasing and decreasing $\bar{\kappa}$, 
 we observe hysteresis loops in the eigenvalues as well as in 
the Polyakov loop, as shown in Fig.3. 
Such observation also bears evidence of the first order transition.

Near the critical point of the first order transition, it is always
 difficult to find a global minimum of the free energy among nearly
  degenerate local minima.  To find a true minimum,
  we divide the lattice volume into domains so that we 
   can put different phases obtained by  the hysteresis loop into
    different domains.
 Starting from such a mixed configuration, a state with the highest pressure
pushes away the other metastable states after many update steps.
 We call this as `boundary-shift test'.
In this way, we can determine the most favorable state for each value of
$\bar{\kappa}$ and can estimate $\bar{\kappa}_{\rm c}$.

\begin{figure}[t]
\begin{center}
\includegraphics[width=7.cm,clip=yes]{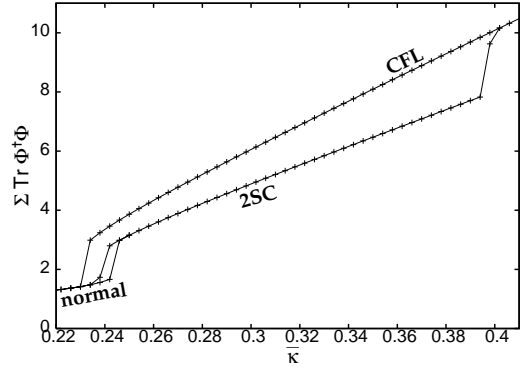} \label{hys}
\vskip -0.3cm
\caption{The hysteresis loop of $\sum_x{\rm Tr}\varPhi^\dag_x\varPhi_x$
obtained by increase or decrease of $\bar\kappa$.
$(\bar\lambda_1,\bar\lambda_2)=(0.01,0.005)$
and $\beta=5.1$.}
\vskip -0.6cm
\end{center}
\end{figure}

 In Fig.4, we show the phase diagram in the $(\bar{\lambda}_1,\bar{\lambda}_2$)-plane obtained by our lattice simulations. The points marked by the symbols (star, cross,
  square, and plus) correspond to the places of actual numerical simulations.
We find three types of thermal transitions depending on the coupling parameters:
 for large values of $\bar{\lambda}_2$, transition from the normal
phase to CFL takes place as star and cross symbols show.
For small or negative values of $\bar{\lambda}_2$,
transition from the normal phase to 2SC takes place as  
plus symbols show.  In between the two cases, there is a 
parameter region where successive transition, 
normal$\rightarrow$2SC$\rightarrow$CFL, takes place as square symbols show. 

\begin{figure}[b!]
\begin{center}
\begin{minipage}{.45\textwidth}
\hspace{-1.5em}
\includegraphics[width=7.5cm,clip=yes]{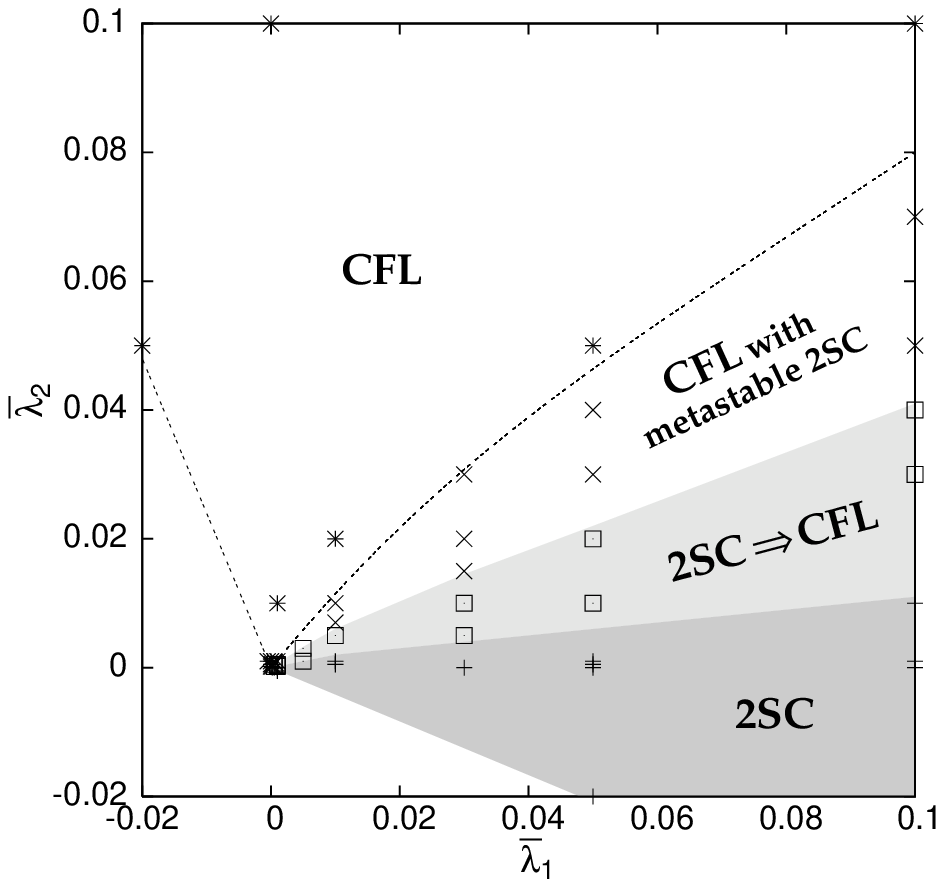} \label{lat}
\vskip -0.6cm
\caption{Phase diagram in the coupling constant space obtained by the lattice
simulation. Symbols show the parameters at which the actual simulations are
 carried out.}
\end{minipage}
\hspace{1.5em}
\begin{minipage}{.45\textwidth}
\includegraphics[width=6.3cm, angle=270,clip=yes]{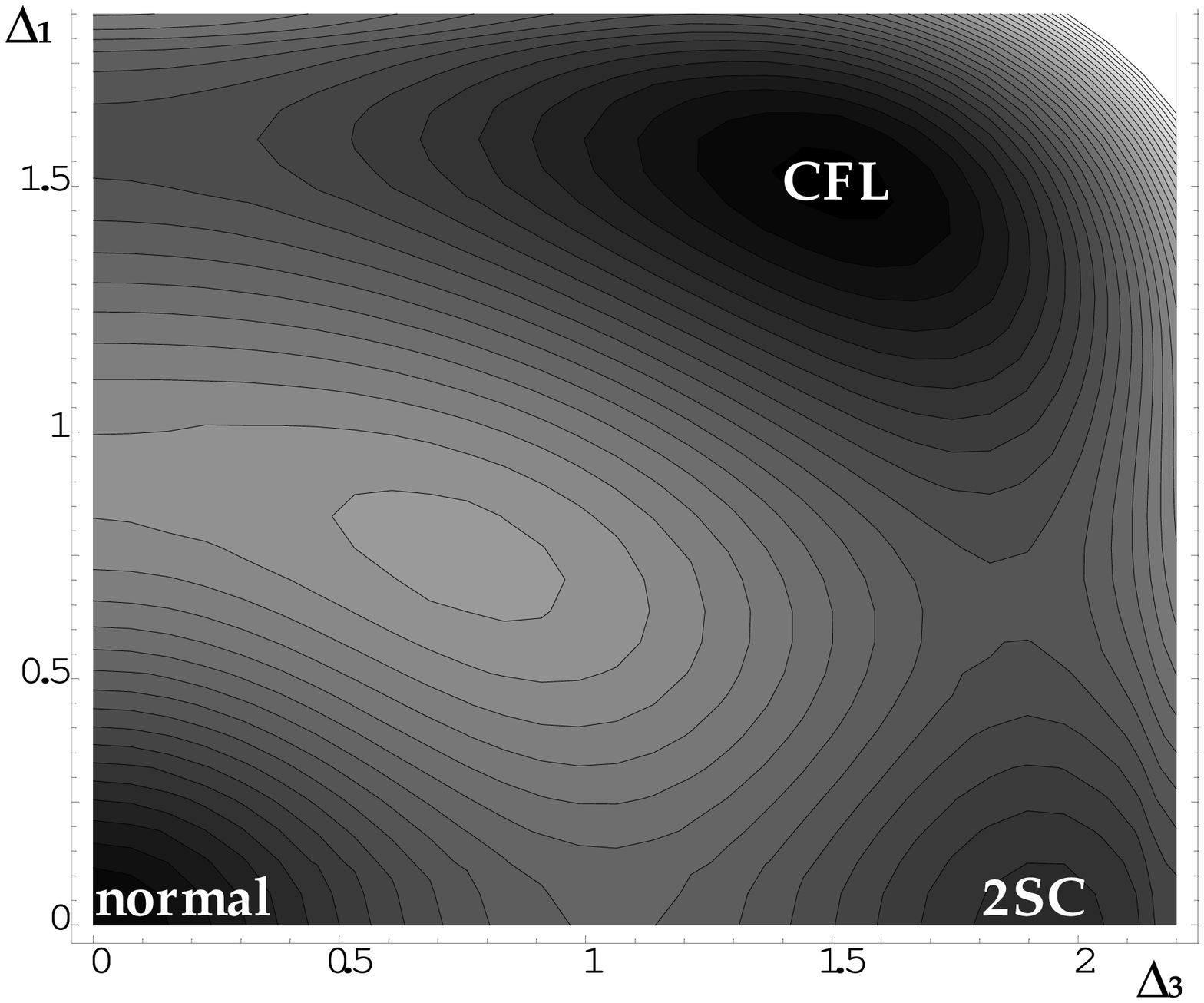} \label{fe}
\vskip -0.3cm
\caption{Contour plot of the free energy at the transition point
from normal phase to CFL in the weak coupling analysis. Scales are in an
arbitrary unit.}
\end{minipage}
\vskip -0.6cm
\end{center}
\end{figure}
 
 The global feature of the phase diagram in Fig.4
is qualitatively similar to that of the weak-coupling analysis shown
in Fig.1b, although the coupling constants 
(e.g. $\lambda_i$ vs. $\bar{\lambda}_i$)
are related in a nontrivial manner.  
Note that, near the phase boundary between 2SC and CFL, 
we find a region (shown by the cross symbols)
where  metastable 2SC survives, even when 
the global minimum of the free energy transfers directly
from the normal phase to CFL.  
This metastable 2SC state, which is 
a local minimum of the free energy, can be seen in 
the hysteresis loop but is pushed away by the `boundary-shift test'. 
The jump of the operators at the transition points
becomes smaller if we take larger $\bar\lambda_i$. 
It is considered that thermal fluctuations of the diquark field
weaken the first order transition.
                                
 Fig.5 shows the contour plot of the free energy at a critical 
 point of the transition from the normal phase to the CFL phase obtained
 in the weak-coupling approach. 
 Setting the diquark field as 
diag$(\Delta_1,\Delta_1,\Delta_3)$, 
the gluonic fluctuations are calculated as
a function of $\Delta_1$ and $\Delta_3$. 
The dark regions have smaller free energy than the light regions.
 The normal, 2SC and CFL phases are identified as 
$\Delta_1=\Delta_3=0$, $(\Delta_1=0,\Delta_3\neq 0)$ and 
$\Delta_1=\Delta_3\neq 0$, respectively. 
 This figure indicates a subtle interplay among three local
 minima as a function of temperature and coupling constants. It also 
serves as an possible explanation of the phase structure shown in Fig.4.

\section{Discussion}
 We have studied thermal transitions of color superconductivity
on the basis of the lattice simulations 
of the Ginzburg-Landau effective action.
 The relevant degrees of freedom
are the diquark field $\varPhi$ and the 
gluon field.
We have identified different  phases of color superconductivity
 by measuring the gauge 
invariant flavor-matrix $\sum_x\varPhi^\dag_x\varPhi_x$
  and extracting its eigenvalues.
 2SC and CFL phases as well as normal phase are clearly 
  distinguished in a gauge invariant way.

 In our simulations, we observed the hysteresis loop of 
various gauge invariant operators as a function of the parameter
 $\bar{\kappa}$,  which designates the first-order thermal transition.
The global minimum of the free energy for each $\bar{\kappa}$
is determined by the boundary-shift method in which 
 we carry out updates by starting from a  mixed configuration obtained 
in the hysteresis loop.
With these procedures, we obtained the phase diagram in the 
coupling constant space and compared the result with that of 
 the weak coupling calculation. 
Three types of thermal transitions are observed
depending on the couplings: normal$\rightarrow$2SC, 
 normal$\rightarrow$CFL and a novel transition,
 normal$\rightarrow$2SC$\rightarrow$CFL.

 To find a realistic phase diagram of color superconductivity
  in  the temperature ($T$) and chemical potential ($\mu$) plane,
 it is necessary to make a connection
  of our coupling constants to $T$ and $\mu$, which requires
  a matching of the GL effective theory to the  microscopic QCD. 
 The quark masses and  color and charge neutralities
  become important at low baryon density and will lead to
  more complex phase structure than that shown in this report.
 Further studies are needed to attack these problems.

\section*{Acknowledgments}
We would like to thank M.~Tachibana for helpful discussions.
We are grateful to T.~Matsuura for valuable comments on
the weak coupling analysis.
We would like to thank S.~Datta for providing us with
SU(3) code. Lattice simulations in this work 
were partly done by RIKEN Super Combined Cluster system.
S.D.~is supported by the JSPS Postdoctoral Fellowship
for Foreign Researchers.
T.H.~is supported by Grants-in-Aid of MEXT, No.~15540254.

\end{document}